# The Spin Dynamics of the Spin Ladder, Dimer Chain Material $Sr_{14}Cu_{24}O_{41}$.

Roger S Eccleston[1], Masatomo Uehara[2], Jun Akimitsu[2], Hiroshi Eisaki[3], Naoki Motoyama[3] and Shin-ichi Uchida[3].

[1]*ISIS Facility, CLRC Rutherford Appleton Laboratory, Chilton, Didcot, Oxfordshire OX11 0QX, UK.*
[2]*Department of Physics, Aoyama-Gakuin University, Setagaya-ku, Tokyo, 157, Japan.*
[3]*Department of Superconductivity, The University of Tokyo, Bunko-ku, Tokyo, 113, Japan*

We have performed inelastic neutron scattering on a single crystal sample of $Sr_{14}Cu_{24}O_{41}$ to study the spin dynamics of the $Cu_2O_3$ spin ladder layers, and $CuO_2$ chains. Data collected with incident energies of 50 meV, 200 meV, 350 meV and 500 meV are best fitted with a dispersion with a spin gap of 32.5±0.1 meV and a maximum of 193.5±2.4 meV, consistent with a coupling along the ladders, $J_\parallel$ = 130 meV and a rung coupling $J_\perp$ =72 meV. We find that excitations with an energy transfer of approximately 11.5 meV can be described solely in terms of a dimer chain with an antiferromagnetic intra-dimer coupling, $J_1$ = 11.2 meV, between next-nearest-neighbour Cu ions and a ferromagnetic inter-dimer coupling, $J_2$ = -1.1 meV. The dimers are separated by two Cu ions providing a periodicity for the dimer chain of five units.

Candidate spin ladder systems have been the subject of considerable research attention since Dagotto et al[1] and Rice et al[2] suggested that the S=1/2 two legged ladder should have a spin-liquid ground state with a spin gap to the lowest triplet excited state. They also concluded that under modest hole doping, the ladder may become superconducting. The existence of a spin gap has been observed experimentally in the ladder material $SrCu_2O_3$[3,4,5].

$Sr_{14}Cu_{24}O_{41}$ was first reported by McCarron et al[6]. The crystal structure comprises adjacent spin ladder layers, identical to the ladder layers in $SrCu_2O_3$, and layers which comprise $CuO_2$ chains. It is of great importance in the study of the spin ladder materials because heavily Ca doped samples have been observed to become superconducting under high pressure[7]. In addition, unlike $SrCu_2O_3$, $Sr_{14}Cu_{24}O_{41}$ can be fabricated under ambient pressure, and it is now possible to grow single crystals of sufficient volume for inelastic neutron scattering experiments. The formal valence of the Cu is +2.25 i.e. the material is inherently hole doped. The ladders are formed by 180⁰ Cu-O-Cu bonds. While the ladders are physically close to each other, they are linked by 90⁰ Cu-O-Cu which is a super-exchange path that leads to weak ferromagnetic coupling. The triangular arrangement of Cu ions on adjacent ladders coupled in this way leads to frustration, effectively de-coupling the ladders. In the $CuO_2$ layers the copper ions are also linked via 90⁰ Cu-O-Cu bonds.

Inelastic neutron scattering measurements on single crystal samples provides a unique method of determining the excitation spectrum for comparison with the excitation spectrum determined theoretically. Previous inelastic neutron scattering measurements have been performed either on polycrystalline samples of Ca doped $Sr_{14}Cu_{24}O_{41}$, or on a steady state source where measurements were restricted to energy transfers below 15 meV. In their study of a polycrystalline sample of $(Sr_{0.8}Ca_{0.2})_{14}Cu_{24}O_{41}$ Eccleston et al[8] observed a spin gap of 34.9 meV for the spin ladders and excitations arising from a dimerised chain centred on 10 meV. Matsuda et al[9] studied single crystals of $Sr_{14}Cu_{24}O_{41}$ on a triple-axis spectrometer. They observed scattering between 9 and 14 meV which they attribute to the dimerized state of the $CuO_2$ chain, with the dimers separated by 2 and 4 times the distance between the nearest neighbour copper ions in the chain. They also observed scattering at approximately 11 meV which they interpreted as arising from a dimerized state in the ladders formed between nearest-neighbour copper ions which are connected by the inter-ladder coupling.

In this paper we present neutron scattering data collected from an array of single crystals of $Sr_{14}Cu_{24}O_{41}$. As far as we are aware, this is the first single crystal inelastic neutron scattering study of a cuprate ladder material which has been able to access the excitations from the spin ladders. Our data show a spin gap for the ladder excitations of 32.5 ± 0.1 meV, with the scattering function, $S(Q,\omega)$, very sharply peaked about the antiferromagnetic zone centre. Fits to our data yield a spin wave maximum of 193.5 ± 2.4 meV although the low intensity of the signal away from the antiferromagnetic zone centre means that this value is not well defined. Like Matsuda et al, we observe scattering in the region of 11 meV, however, our analysis suggests that this scattering may be interpreted in terms of dimerization of the chain, with dimers formed between next-nearest neighbour copper ions which are then ferromagnetically coupled with a period of 5 copper-copper distances.

All data were collected using the High Energy Chopper Spectrometer (HET) on the ISIS Pulsed Neutron Source at the Rutherford Appleton Laboratory in Oxfordshire, UK. The neutron beam is monochromated by a rotating Fermi chopper which is phased to the source, allowing the selection of incident energies between 15 meV and 2 eV. Detectors at 4m cover the angular range for 2.7° to 9° and detectors at 2.5m cover the range from 9° to 30°. Two further detector banks are available at high angles, but were not used in these measurements.

The crystals were grown by the travelling solvent floating zone method. They were identified as $Sr_{14}Cu_{24}O_{41}$ single crystals by X-ray diffraction pattern and Laue method. The typical size of the crystals is about 5mm, 4mm and 50mm parallel to the a, b and c axes, respectively.

An array of 8 mutually aligned single crystals were used, providing a total sample mass of approximately 40g. The crystals were mounted on a closed cycle refrigerator with the c and a axes in the equatorial plane. All data described hereafter were collected with the sample at a temperature of 20K and with $k_i$ perpendicular to c, ie the ladder direction. In this configuration we make full use of the one dimensional nature of the sample and using the full detector array out to 30° simultaneously access a wide portion of $(Q,\omega)$ space. We are assuming that there is no dispersion along the a direction, since any time-of-flight scan in this configuration will have a component in $Q_\perp$ (Q perpendicular to the ladder direction, in the a,c plane) and $Q_\parallel$ (Q parallel to the ladder direction). The magnetic signal is collected in the detectors in the equatorial plane. The background signal arises from single and multi-phonon scattering, and with increasing incident energy, the portion of the background which is multi-phonon scattering will increase. We have used data collected in the vertical banks of HET as a background signal. Measurements performed at room temperature, and with the crystal rotated slightly off axes has shown this to be a valid approximation.

In the first portion of this paper we discuss measurements of the ladder dispersion. In the second section we discuss the lower energy transfer excitations which, we believe, originate in the $CuO_2$ chains.

Data were collected at with incident energies of 50meV, 80meV, 200meV, 350meV and 500meV. Bright spots of intensity were observed at Q=0.8Å$^{-1}$ and weaker spots at 2.4Å$^{-1}$, which are the π and 3π points respectively, but scattering away from these points was weak. We chose to sum the data collected in the 4m, horizontal detector banks at each incident

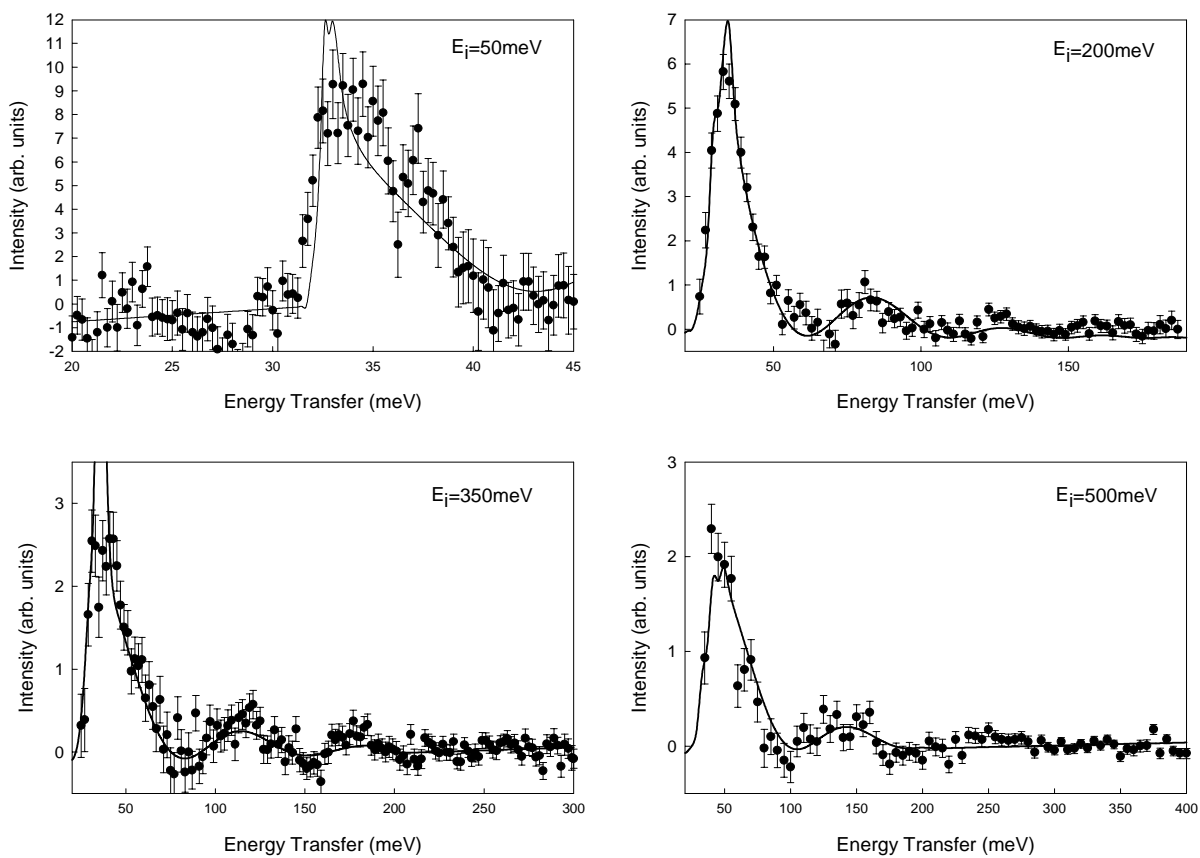

Figure 1. Scattering summed over the low angle 4m bank on HET with incident energies 50, 200, 350 and 500meV with the results of the fits described in the text.

energy to improve statistics.

We chose a nominal one dimensional dispersion relation, and calculated the integration (effective magnetic density of states, g(ω)) corresponding to the summation of the scattering anticipated form this dispersion over the region of (Q,ω) space over which data was collected. We then multiplied this function by the projection of $S(Q_\perp,\omega)$ on the measured trajectory, a projection of a nominal form for $S(Q_\parallel,\omega)$ on the trajectory of the measurement, the square of the magnetic form factor, $F^2(Q)$, and the temperature factor, $T(\omega)$. This function was then convoluted with the instrumental resolution function, determined from measurements on a vanadium standard sample, and fitted to the data collected with incident energies of 200, 350 and 500 meV simultaneously using the Multi-Frills software[10]. Data collected with an incident energy of 50meV were independently fitted in the same manner.

The dispersion relation we have chosen for convenience is

$$\hbar\omega(q)^2 = E_g^2 + C^2 \sin^2(q-\pi) \quad (1)$$

where $E_g$ is the spin gap and C is the band maximum. While this dispersion differs the one predicted for a spin ladder, they are very similar over the range of (Q,ω) space accessed in this experiment, and this expression has the benefit of making the integration straightforward.

Since the time of flight scan has a component in both $Q_\perp$ and $Q_\parallel$ we must make allowance for the modulation in the structure factor parallel to the rungs. Tennant[11] has shown that $S(Q_\perp)$ is given by

$$S(Q_\perp) \propto 1-\cos(q_\perp b) \quad (2)$$

for k=π ie single magnon excitations.

Again for simplicity, and in the absence of a more valid form for $S(Q_\parallel,\omega)$ we have chosen an exponential form, thus.

$$S(Q_\parallel,\omega)=e^{\beta\hbar\omega} \quad (3)$$

Thus the overall expression used to fit the data can be summarised as

$$S(Q_\parallel,Q_\perp,\omega)=g(\omega)S(Q_\perp,\omega)S(Q_\parallel,\omega),T(\omega)F^2(Q). \quad (4)$$

The fits to the four data sets are shown in figure 1, and yielded the values C = 193.5 ± 2.4 meV, $E_g$ = 32.5 ± 0.1 meV and β = -0.011 ± 0.001. From figure 5 of Barnes and Riera[12] it is clear that the ratio of $E_g$ to C is a function of the ratio of the rung coupling to the chain coupling, $\alpha = J_\perp/J_\parallel$, and we are able to extract the following relation from that figure, which yields α = 0.55 for our parameters.

$$\frac{E_g}{C} = 0.08 + 0.16(\alpha) \quad (5)$$

Similarly, the relationship between $E_g/J$ and α, again based on Barnes and Riera's calculations, is given by Johnston[13].

$$\frac{E_g}{J} = 0.4(\alpha) + 0.1(\alpha)^2 \quad (6)$$

Thus, our parameters yield values of $J_\parallel$ = 130 meV and $J_\perp$ = 72 meV.

We now turn our attention to the lower energy excitations. By integrating the data collected with an incident energy of 50 meV between ℏω=9 meV and ℏω=14 meV we have extracted the Q dependence of the scattering (figure 2), and fitted it to the structure factor for a dimer chain[14]:

$$S(Q) = A(1-\cos(q.d))\left(1+\frac{\alpha_c}{2}\cos(q.b)\right)$$
$$|F(Q)|^2 + c \quad (7)$$
$$\alpha_c = \frac{J_2}{J_1}$$

where d is the intra-dimer distance, b is the inter-dimer distance, c is a linear background and A is a constant scaling factor. The fit yields a value of d=5.48 ± 0.06 Å., b=14.1 ± 1.1 Å and $\alpha_c$=-0.1 ± 0.1. Clearly the last two parameters are not well defined from this fit, however, the Cu-Cu distance in the chain is 2.75 Å, so it is reasonable to suggest that dimers are formed between some next-nearest-neighbour Cu ions.

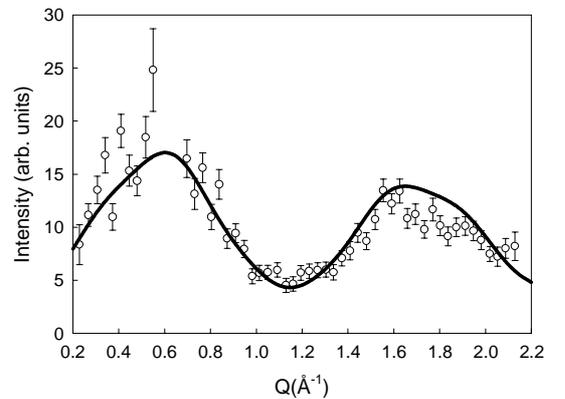

Figure 2 Scattering data collected with an incident energy of 50meV integrated between hw=14 meV and 9 meV. The line represents a fit to equation 7.

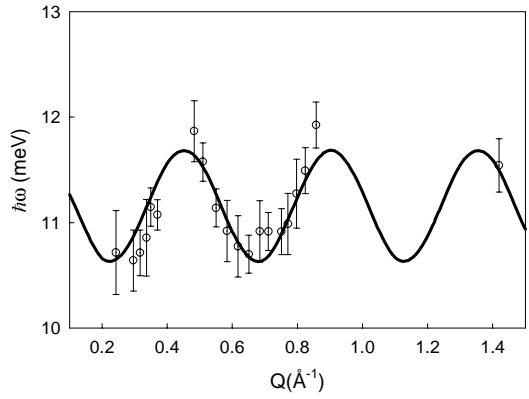

Figure 3. Dispersion relation taken from the data collected with an incident energy of 30 meV. The line represents a fit to equation 8.

Fitting the same expression to data collected at 30meV yields a slightly smaller result of 5.36 ± 0.08 Å. The data collected at 30meV (figure 3) shows dispersion which can be fitted to the dispersion for a dimer chain[15]

$$\frac{\hbar\omega(q)}{2JS} = \left(1 - \frac{1}{16}\alpha_c^2\right) - \left(\frac{1}{2}\alpha_c + \frac{1}{2}\alpha_c^2\right)\cos(q.b) \\ - \left(\frac{1}{16}\alpha_c^2 \cos(2q.b)\right) \quad (8)$$

yielding a value for the inter-dimer distance, b, of 13.9±0.2 Å, approximately equivalent to five copper-copper distances along the chain, and values for the intra- and inter-dimer exchanges of $J_1$ = 11.18±0.05 meV and $J_2$ = -1.12±0.16 meV i.e the inter-dimer coupling is weak and ferromagnetic. Hiroi et al[16] have pointed out that the dimerization in the $CuO_2$ chains may be caused by the localisation of holes at low temperatures on the structurally modulated chains. They suggest that the most probable deformation of the $CuO_2$ chains has a period which is commensurate with ten copper-copper distances. The dimerization pattern suggested by our data is also consistent with the six holes per formula unit forming Zhang-Rice singlets on the chains rather than on the ladders to produce the non-magnetic separators between the magnetic ions which make up the dimerized chain.

We have studied the excitations from the spin ladders and spin chains in a single crystal array of $Sr_{14}Cu_{24}O_{41}$. We have observed a spin gap of 32.5 ± 0.1 meV and a band maximum of 193.5± 2.4 meV which provides values of $J_\parallel$ = 130 meV and $J_\perp$ = 72 meV. The scattering is sharply peaked around the antiferromagnetic zone centre. There is no scattering arising from band minima at either the 0 or $2\pi$ points in the Brillouin zone, consistent with anticipated forms for the spin ladder dispersion[17,18]. Excitations measured at energy transfers of 15 meV and below are consistent with a chain made up of dimers of next-nearest neighbour copper ions coupled antiferromagnetically which are separated by three non-magnetic $Cu^{3+}$ ions and are weakly, ferromagnetically coupled.

We are grateful to Professor Ted Barnes and Dr David Johnston for several very useful discussions. This work was partially funded by the UK-Japan collaboration on neutron scattering.